# Statistical properties of fidelity in quantum tomography protocols in Hilbert spaces of different dimensions

Yu. I. Bogdanov[1*], I.D. Bukeev[1,2], A. K. Gavrichenko[1]

[1] *Institute of Physics and Technology, Russian Academy of Sciences, Moscow, 117218 Russia*
[2] *Moscow Institute of Physics and Technology, Dolgoprudny, Moscow Region, 141700 Russia*

**Abstract** - A throughout study of statistical characteristics of fidelity in different protocols of quantum tomography is given. We consider protocols based on geometry of platonic solids and other polyhedrons with high degree of symmetry such as fullerene and its dual polyhedron. Characteristics of fidelity in different protocols are compared to the theoretical level of the minimum possible level of fidelity loss. Tomography of pure and mixed states in Hilbert spaces of different dimension is analyzed. Results of this work could be used for a better control of quantum gates and quantum states in quantum information technologies.

## 1. Introduction

Quantum information technologies rely on the use of quantum states in novel data transmission and computing protocols [1–4]. Control is achieved by statistical methods via quantum state reconstruction. At present quantum state and process tomography serves as a principal instrument for characterization of quantum states preparation and transformation quality [5-31].

In this paper, we use a new methodology for statistical reconstruction of quantum states which is based on analysis of completeness, adequacy and accuracy of quantum measurement protocols [32-35]. Completeness is assessed by means of singular value decomposition of a matrix built upon operators of measurement. Adequacy of a protocol implies redundancy of a measurement protocol compared to the minimum number of measurements required for reconstruction. Adequacy is assessed by consistency of redundant statistical data and quantum theory. Accuracy of statistical reconstruction of quantum states is studied by a universal statistical distribution proposed in [34].

Multi-qubit protocols described in this work are formed by projection quantum measurements on states that are tensor products of single-qubit states. If a single-qubit state is formed by a polyhedron with $m$ faces and therefore has $m$ rows then the $l$ qubit protocol which corresponds to it will have $m^l$ rows.

In the basis of single-qubit protocols it is worth highlighting regular polyhedrons and polyhedrons with lesser but still rather high level of symmetry.

Regular polyhedrons (Platonic solids) are used for the most symmetrical and uniform distribution of quantum states on the Bloch sphere. States that define quantum states projection are defined by directions from the centre of Bloch sphere to centers of polyhedron faces. Therefore, the number of polyhedron's faces defines the number of protocol's rows and is equal to 4 for tetrahedron, 6 for cube, 8 for octahedron, 12 for dodecahedron and 20 for icosahedron.

These five bodies form the complete set of regular polyhedrons. Search for protocols of quantum measurements on Bloch sphere with high symmetry and number of rows greater that twenty force us to consider non-regular polyhedrons that have high symmetry. As examples of such polyhedrons we have chosen fullerene (truncated icosahedron) that defines quantum measurement protocol with 32 rows (equal to the number of fullerene's faces) and also a dual to fullerene polyhedron (pentakis dodecahedron) which defines quantum measurement protocol with 60 rows (which is the number of its faces and also the number of vertices of fullerene).

---

[*] e-mail: bogdanov@ftian.ru

It is noteworthy that all protocols considered here can be brought to decomposition of the unity [36].

Comparison of the maximum possible fidelity with fidelity of protocols considered here shows that as the number of polyhedrons' faces increases fidelity rapidly converges to the theoretical limit (in addition, rapidly increases a uniformity of fidelity distribution on the Bloch sphere). We should mention that accuracy of suggested protocols is much higher in comparison with an accuracy which provide not so highly symmetrical protocols.

Considered method is generalized on the case of multi-qubit state tomography and accepts the reconstruction of not only pure states but mixed states of arbitrary rank too. Developed method is addressed to increase the accuracy and efficiency of quantum tomography procedures.

## 2. Precision of quantum tomography

Precision of quantum tomography can be defined by a parameter called Fidelity [1,37]

$$F = \left(Tr\sqrt{\rho_0^{1/2} \rho \rho_0^{1/2}}\right)^2, \quad (1)$$

where $\rho_0$ is theoretical density matrix and $\rho$ is reconstructed density matrix.

Fidelity shows how close the reconstructed state is to the ideal theoretical state. The reconstruction is precise if Fidelity is equal to one.

This equation looks quite complex, but it becomes simple if we apply the Uhlmann theorem [37]. According to the theorem, Fidelity is simply the maximum possible squared absolute value of the scalar product, which we may obtain using purification procedure.

$$F = |\langle c_0 | c \rangle|^2, \quad (2)$$

where $c_0$ and $c$ are theoretical and reconstructed purified state vectors.

We explicitly use the Uhlmann theorem in our algorithm of statistical reconstruction of quantum states. This fact is very important. Even if the state is not pure, we have to purify it by moving into the space of higher dimension [34].

It is well known that purified state vectors are defined ambiguously. However, this ambiguity does not preclude us from reconstructing a quantum state. This is a very useful feature of our algorithm. It is devised in the way that different purified state vectors produce the same density matrix and therefore the same fidelity during the reconstruction. This principle is very important for proposed procedure and thus reconstruction can be held by means of purification. Purification greatly facilitates the search of solution, especially when we need to estimate a great number of parameters (hundreds or even thousands).

## 3. Generalized statistical distribution in the problem of quantum state reconstruction

It is equally important that due to the usage of purification procedure we succeed in formulating a generalized statistical distribution for fidelity [34]. The value $1-F$ can be called the loss of fidelity. It is a random value and its asymptotical distribution can be presented in the following form:

$$1 - F = \sum_{j=1}^{j_{\max}} d_j \xi_j^2 \quad (3),$$

where $d_j \geq 0$ are non-negative coefficients, , $\xi_j \sim N(0,1)$ $j = 1,...,j_{\max}$ are independent normally distributed random values with zero mean and variance equal to one, $j_{\max} = (2s - r)r - 1$ is the number of degrees of freedom of a quantum state and corresponding distribution; $s$ is the Hilbert space dimension, $r$ is the rank of mixed state, which is the number of non-zero eigenvalues of the density matrix. In particular $j_{max} = 2s - 2$ for pure states and $j_{max} = s^2 - 1$ for mixed states of full rank ($r = s$).

This distribution is a natural generalization of chi-squared distribution. Ordinary Chi-squared distribution corresponds to the particular case when $d_1 = d_2 = ... = d_{j_{max}} = 1$ (all components of vector $d$ are equal to one).

From (3) we get that average fidelity loss is equal to this expression.

$$\langle 1 - F \rangle = \sum_{j=1}^{j_{max}} d_j \quad (4)$$

It is also easy to show that the variance for fidelity loss is given by the following equation:

$$\sigma^2 = 2 \sum_{j=1}^{j_{max}} d_j^2 \quad (5)$$

We can analytically calculate moments of higher order for this distribution. For example the momentum of third order is called skewness and describes asymmetry $\beta_1$. The fourth-order moment defines the value called excess kurtosis. $\beta_2$ The equations for the values are as follows:

$$\beta_1 = \frac{8 \sum_{j=1}^{j_{max}} d_j^3}{\sigma^3} \quad (6)$$

$$\beta_2 = \frac{48 \sum_{j=1}^{j_{max}} d_j^4}{\sigma^4} \quad (7)$$

Recall that for a random variable $x$ characteristics are by definition:

$$\beta_1 = \frac{M[(x - M(x))^3]}{\sigma^3} \quad (8)$$

$$\beta_2 = \frac{M[(x - M(x))^4]}{\sigma^4} - 3, \quad (9)$$

where $M$ denotes mathematical expectation.

In the asymptotical limit considered by us, parameters $d_j$ are inversely proportionate to the sample size $n$. Let us introduce the value of fidelity loss which is independent from sample size.

$$L = n \langle 1 - F \rangle = n \sum_{j=1}^{j_{max}} d_j \quad (10)$$

This quantity will be the main characteristic of precision in examples mentioned later. However, prior to examples it is better to study what the protocol of measurement is.

## 4. How does a quantum measurement protocol works

A quantum measurement protocol can be defined by a so-called instrumental matrix $X$ that has $m$ rows and $s$ columns [16-18]. Here $s$ is a number of Hilbert space dimensions; $m$ is the number of projections in such space. For every row, i.e. for every projection, there is corresponding amplitude $M$.

$$M_j = X_{jl} c_l \quad j = 1, 2, ..., m \quad (11)$$

Here we assume a summation by the joint index $l$.

The square of the absolute value of the amplitude defines the intensity of a process $\lambda$, which is the number of events in one second.

$$\lambda_j = |M_j|^2 \quad (12)$$

The number of registered events $k_j$ is a random variable that has Poisson distribution. Lambda is the parameter in Poisson distribution; $t_j$ is the time of exposition of the selected row of the protocol.

$$P(k_j) = \frac{(\lambda_j t_j)^{k_j}}{k_j!} \exp(-\lambda_j t_j) \quad (13)$$

It is convenient to introduce special observables - so called intensity operators. These observables are measured by the protocol during experiment.

$$\lambda_j = tr(\Lambda_j \rho) \quad (14)$$

Here $\Lambda_j = X_j^+ X_j$ is intensity operator for quantum process, $X_j$ is the row of the instrumental matrix $X$.

In this case the intensity operator for quantum process $\Lambda_j$ is a projector, so we have

$$\Lambda_j^2 = \Lambda_j \quad (15)$$

Formally, in the more general case, $\Lambda_j$ is arbitrary positively defined operator [36].

It can be presented as a mixture of projection operators described above.

$$\Lambda_j = \sum_k f_k X_j^{(k)+} X_j^{(k)} \quad (16)$$

Here index $k$ sums different components of the mixture that have weights $f_k > 0$.

Such measurement can be conveniently presented as a reduction of the set of projection measurements where only total statistics is available, while statistical data for individual components are not available. The general projection measurement is a particular case of equation (16) where $f_1 = 1$, $f_2 = f_3 = ... = 0$.

If the sum of intensities multiplied by exposition time is proportionate to a unit matrix that we shall say that the protocol is brought to decomposition of unity [36].

$$I = \sum_{j=1}^{m} t_j \Lambda_j = I_0 E \quad (17)$$

where $I_0$ is a constant which defines overall intensity.

In that case the protocol analysis is simplified. However it is worth mentioning that real experimental protocols often cannot be brought to decomposition of unity. Our method is however equally applicable to these cases too.

The normalization condition for the protocol defines the total expected number of events $n$ summarized by all rows:

$$\sum_{j=1}^{m} \lambda_j t_j = n \quad (18)$$

where $t_j$ is the acquisition time.

For any protocol of quantum measurements we can define two important notions - completeness and adequacy.

## 5. Completeness and adequacy of protocol

We define the row of the measurement matrix $B$ for a tomographic protocol as the direct product of row $X_j$ and its complex conjugate row $X_j^*$: $B_j = t_j \cdot X_j^* \otimes X_j$, its size being $m \times s^2$ (we assume $m \geq s^2$). With this matrix $B$, the protocol can be compactly written in the matrix form:

$$B\rho = K \qquad (19)$$

Here $\rho$ is the density matrix, given in the form of a column (second column lies below the first, etc.). The vector $K$ of length $m$ records the total number of registered outcomes. The algorithm for solving equation (19) is based on the so called singular value decomposition (svd) [38]. Svd serves as a base for solving inverse problem by means of pseudo-inverse or Moore-Penrose inverse [38,39]. In summary, matrix $B$ can be decomposed as:

$$B = USV^+ \qquad (20)$$

where $U$ ($m \times m$) and $V$ ($s^2 \times s^2$) are unitary matrices and $S$ ($m \times s^2$) is a diagonal, non-negative matrix, whose diagonal elements are "singular values". Then equation one (19) transforms to a simple diagonal form:

$$Sf = Q \qquad (21)$$

with a new variable $f$ unitary related to $\rho$ via $f = V^+ \rho$ and a new column $Q$ unitary related to the vector $K$ by equation $Q = U^+ K$. We use this algorithm as a zero approximation for maximal likelihood state reconstruction.

By defining $q$ as the number of non-zero singular values of $B$ we formulate two important conditions of any tomography protocol, namely its completeness and adequacy [35]. The protocol is supposed to be informationally complete if the number of tomographically complementary projection measurements is equal to the number of parameters to be estimated; mathematically completeness means $q = s^2$.

Adequacy means that the statistical data directly correspond to the physical density matrix (which has to be normalized, Hermitian and positive). However, generally for mixed state it can be tested only if the protocol consists of redundant measurements, i.e. if $m > s^2$.

## 6. The maximum possible precision

The lower the value of the loss function (10), the higher is the precision of the protocol. As we can see from the equation (10), this value is determined by vector $d$ that defines the general fidelity distribution that derived in [34]. Therefore the general distribution for fidelity allows us to completely solve the problem of precision for quantum tomography.

It appears that the minimum possible loss is given by the following equation:

$$L_{min}^{opt} = \frac{v^2}{4(s-1)} \qquad (22)$$

Here $v$ is the number of parameters that we need to estimate $v = (2s - r)r - 1$, $r$ is the rank of mixed state, $r = 1$ for a pure state, $r = s$ for a mix of full rank.

For pure states $v = 2s - 2$, therefore possible loss is given by the following equation:

$$L_{min}^{opt} = s - 1 \qquad (23)$$

For mixed states of full rank $v = s^2 - 1$, and therefore possible loss is given by the following equation:

$$L_{min}^{opt} = \frac{(s+1)^2 (s-1)}{4} \qquad (24)$$

Any protocol for any quantum state cannot have losses lower that those defined by this equation, if the protocol can be brought to decomposition of unity.

Note that if the protocol cannot be brought to decomposition of unity then losses can be lower than defined by this equation. However this is true only for certain states and the improve in precision of reconstruction for some states is completely compensated by a significant deterioration in reconstruction precision for other states.

# 7. Scanning the Bloch sphere

The results allow for vivid illustrations for the cases of single qubit pure states. The set of pictures (images of polyhedrons are got from http://en.wikipedia.org) that we demonstrate shows the results of the Bloch sphere scanning by means of various measurement protocols. The color describes the value of Fidelity loss function.

The first picture defines the value of the loss function for the protocol based on tetrahedron. Here as well as on the other pictures the minimum losses are equal to 1. The maximum losses for tetrahedron are equal to 3/2.

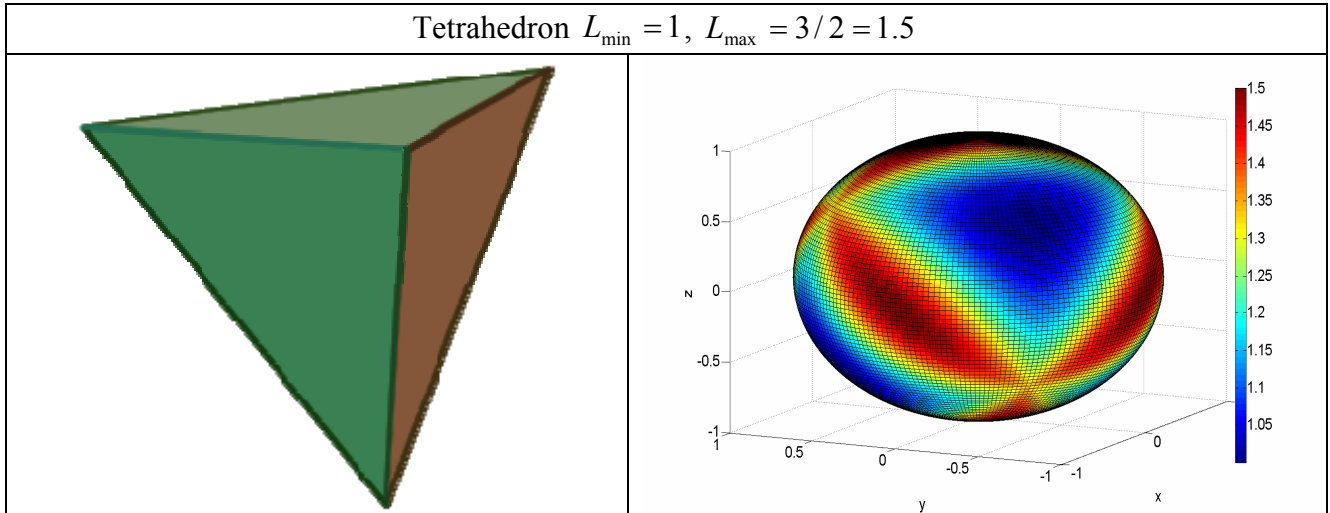

Figure 1. Value of the loss function for tetrahedron

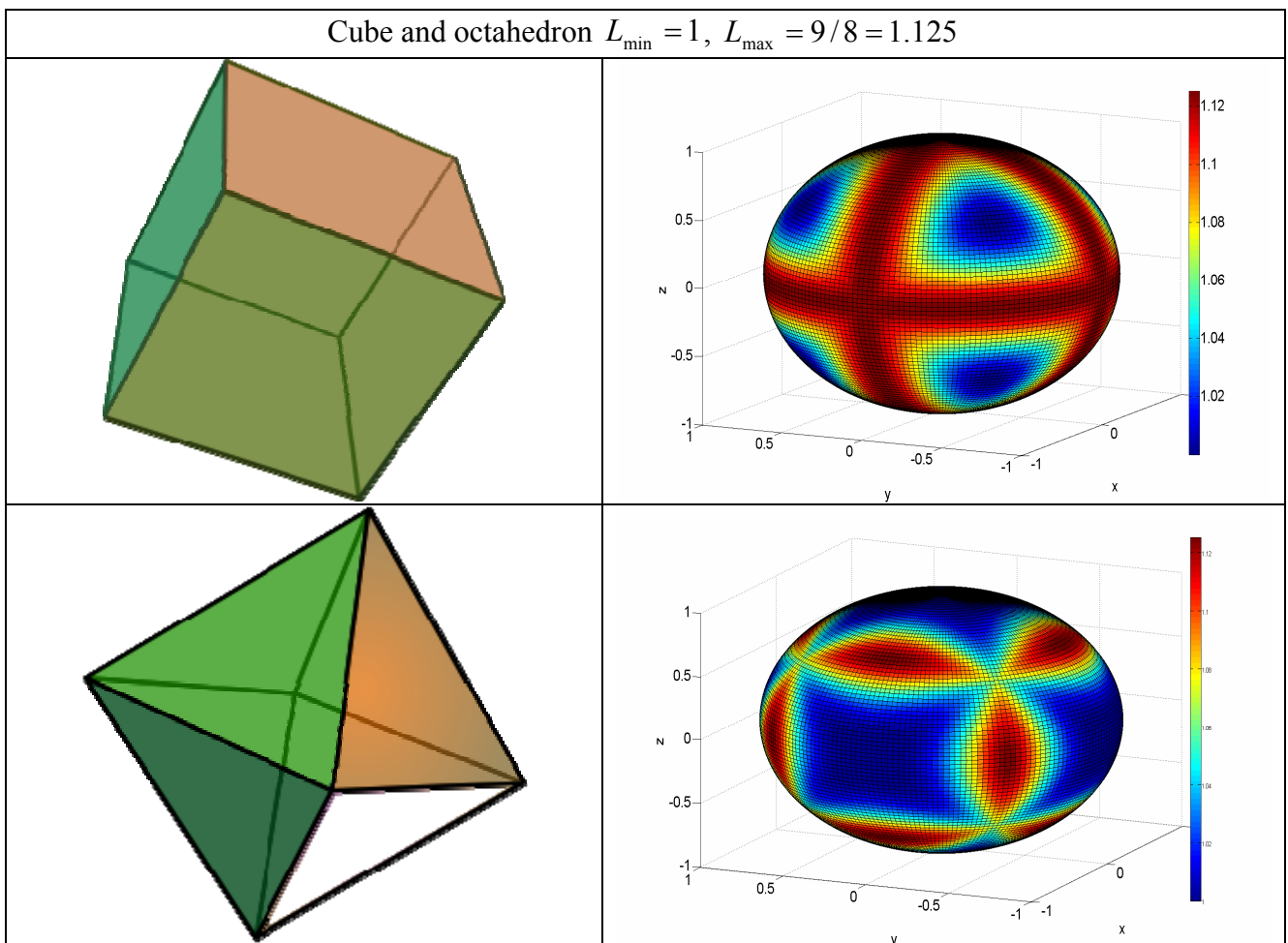

Figure 2. Value of the loss function for cube and octahedron

On the previous picture we present a cube and an octahedron. These polyhedrons are dual to each other. The maximum losses are equal to 9/8 in both cases.

On the next two pictures we present a dodecahedron and an icosahedron, as well as fullerene and a polyhedron that is dual to the latter.

We can see that as the number of projections grows the maximum losses converge to the minimum possible losses. In the limit of infinite number of points on the Bloch sphere we get an optimal protocol for which the precision of reconstruction does not depend on the reconstructed state at all.

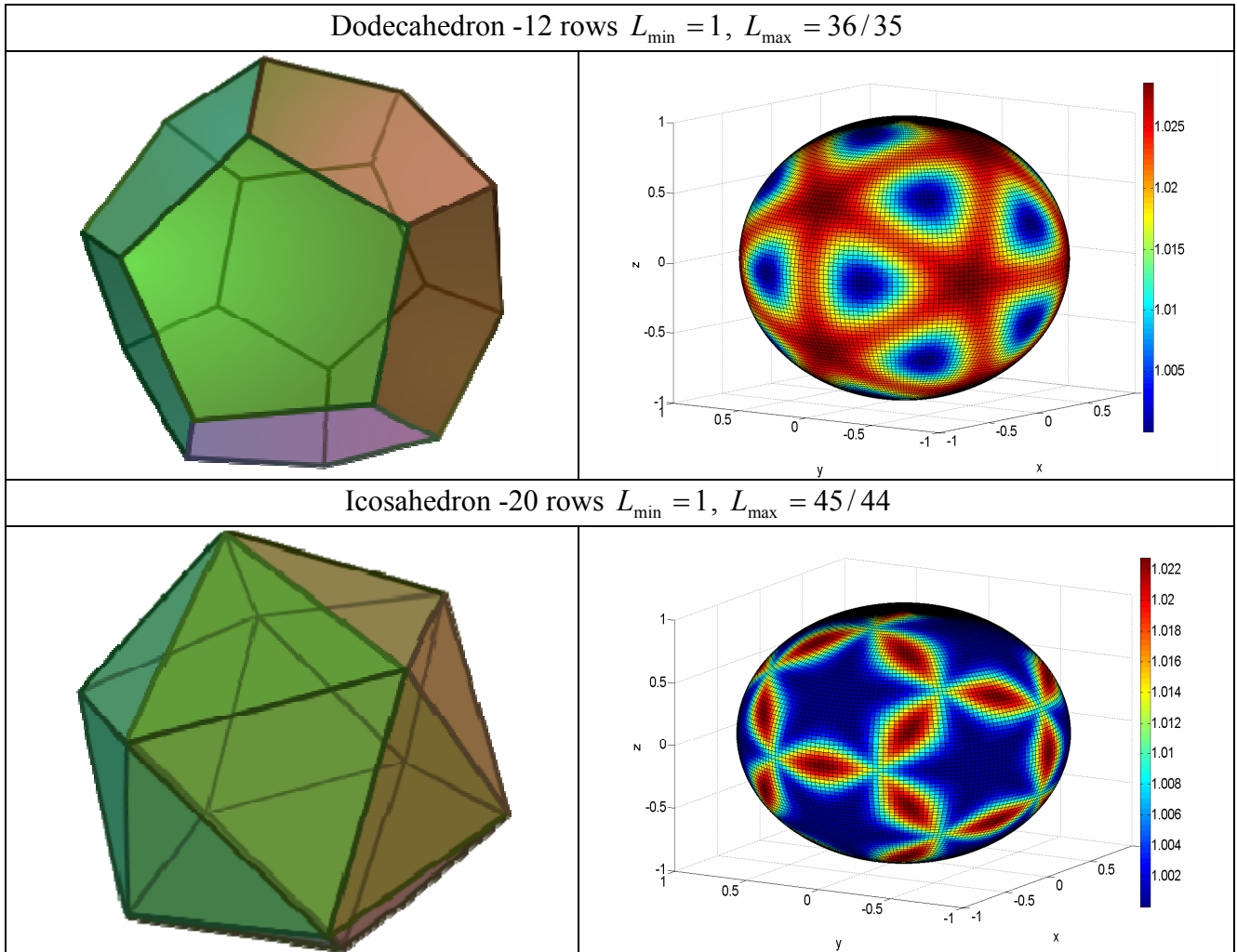

Figure 3. Value of the loss function for dodecahedron and icosahedron

Dodecahedron -12 rows $L_{min} = 1$, $L_{max} = 36/35$

Icosahedron -20 rows $L_{min} = 1$, $L_{max} = 45/44$

And in addition, fullerene and its dual protocols are presented on the last picture.

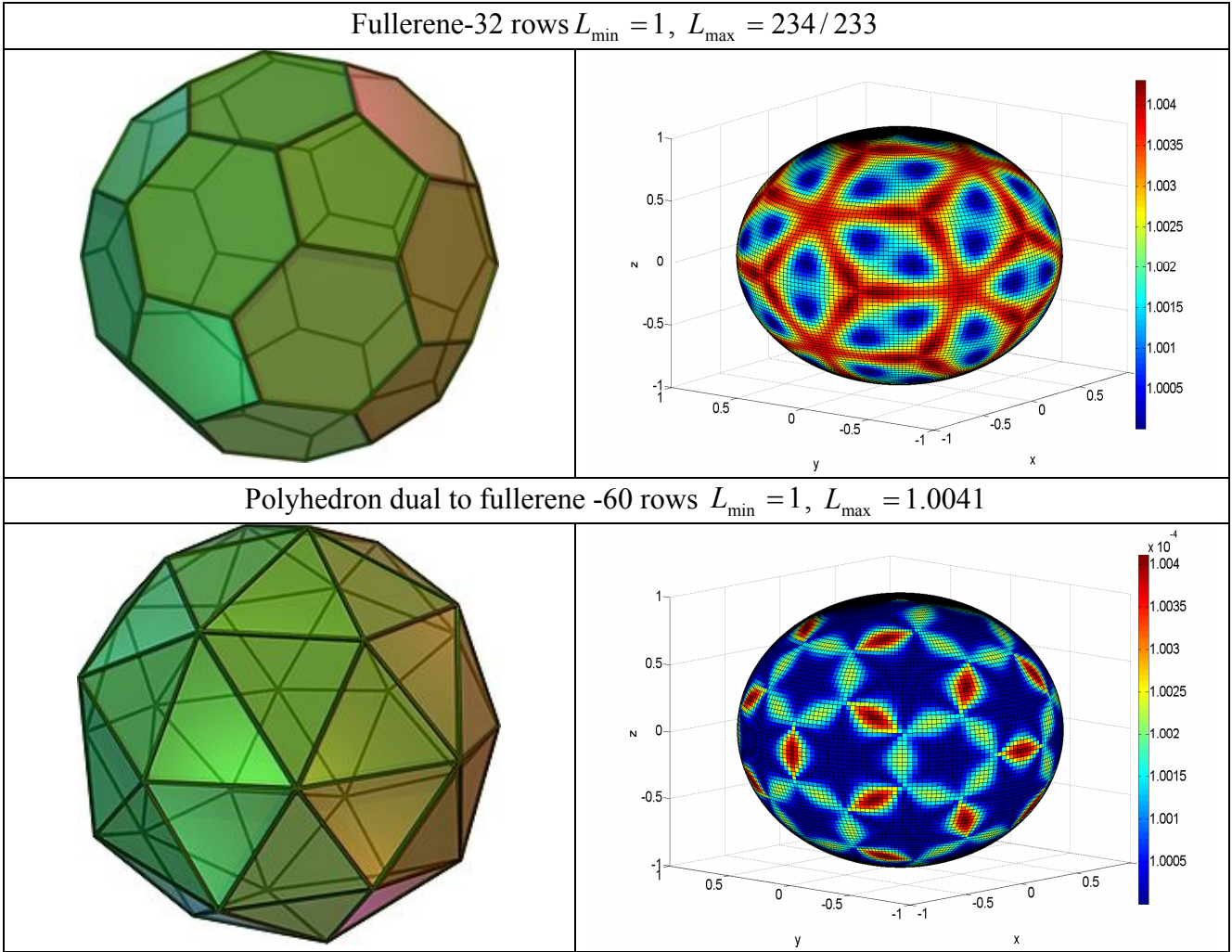

Figure 4. Value of the loss function for fullerene and its dual polyhedron

## 8. Real precision of protocols (for pure states)

Below in the table 1 we present results of numerical experiments for pure quantum states with the number of qubits from 1 to 3.

Precision of every protocol can be characterized by the following relation $L_{min} \leq L \leq L_{max}$. This inequality defines a rather narrow interval, where the precision of quantum state reconstruction lies definitely.

Numerical calculations demonstrate that minimum possible losses $L_{min}$ are defined by the theoretically derived optimal limit $L_{min} = L_{min}^{opt} = s - 1$.

Let we discuss the upper limit $L_{max}$ shown in the table as the result of numerical experiments. We can see that for single qubit protocols as the number of projections grows the maximum losses converge to the minimum possible losses. In the limit of infinite number of points on the Bloch sphere we get an optimal protocol for which the precision of reconstruction does not depend on the reconstructed state.

It is not true for multi-qubit protocols. For two-qubit protocols, the maximum possible losses approach the level $L_{max} \approx 3.37 - 3.38$ while the minimum possible level is equal to $L_{min}^{opt} = 3$.

Similarly for three-qubit states the values $L_{max} \approx 7.7$ and $L_{min}^{opt} = 7$. These results are due to the fact that the protocols perform projections only on non-entangled states.

Table 1. Results of numerical experiments that define maximum precision losses.

|  | 1 qubit ($s = 2$ $L_{min} = 1$) | 2 qubits ($s = 4$ $L_{min} = 3$) | 3 qubits ($s = 8$ $L_{min} = 7$) |
| --- | --- | --- | --- |
| Tetrahedron ($m = 4$) | $3/2 = 1.5$ | $4.442971458$ | $\approx 10.4$ |
| Cube ($m = 6$) | $9/8 = 1.125$ | $\approx 3.5839$ | $\approx 8.2$ |
| Octahedron ($m = 8$) | $9/8 = 1.125$ | $3.4708(3)$ | $\approx 7.9$ |
| Dodecahedron ($m = 12$) | $36/35$ | $\approx 3.42$ | $\approx 7.8$ |
| Icosahedron ($m = 20$) | $45/44$ | $\approx 3.39$ | $\approx 7.8$ |
| Fullerene ($m = 32$) | $\approx 234/233$ | $\approx 3.38$ | $\approx 7.7$ |
| Polyhedron dual to fullerene ($m = 60$) | $1.004103748\,8$ | $\approx 3.38$ | $\approx 7.7$ |

So from the theoretical point of view in the multi-qubit case our protocols are not the best possible because they do not use projections on entangled states. To avoid confusion it is better to note that the protocols allow one to reconstruct entangled states as well, even though they use only projections on non entangled states. In that case the precision will be somewhat smaller than the minimum possible limit. ($L_{max} \approx 3.37$ compared to $L_{min}^{opt} = 3$ for two-qubit states and $L_{max} \approx 7.7$ compared to $L_{min}^{opt} = 7$ for three-qubit states).

Also worth noting that though the protocols based on polyhedrons with small number of faces (tetrahedron, cube) are somewhat less precise; they are much easier in practical implementation.

## 9. Real precision of protocols (for mixed states)

When considering tomography of mixed states it is worth to note that for mixed states there is no finite upper limit for precision losses (losses can be infinitely large $L \to \infty$). Such large losses are inherent to mixed states which are close to pure ones. In fact the number of real parameters that define a mixed state of full rank in Hilbert space of dimension $s$ is equal to $s^2 - 1$, which is significantly greater for large $s$ than for a pure state that takes only $2s - 2$ real parameters. In case of a mixed state that has one predominant component, the smaller weight components almost do not influence statistical data and do not increase the amount of Fisher information for reconstructing the greatly larger number of parameters. Numerical experiments completely prove these considerations.

At the same time the lower limit for precision losses can be applied to mixed states. In this case optimal minimum losses are realized for "white noise states" (uniform density matrix), when all components have equal weights. Then there is a simple relation between vector $d$ of dimension

$s^2 - 1$ that defines distribution of precision loss and the vector of singular values of measurement matrix $B$ (we need to delete the largest value responsible for normalization from the vector of singular values of dimension $s^2$). Let us denote by $b$ the reduced vector of singular values of dimension $s^2 - 1$. Then the relation between vectors is as follows:

$$d_j = \frac{m^l}{4snb_j^2}, \quad (25)$$

where $m$ is the number of faces of a polyhedron, $l$ is the number of qubits in registry and $n$ is the sample size. The corresponding estimate for $L_{\min}$ for the protocols considered by us is given by the following equation:

$$L_{\min} = \left[ n \sum_j d_j \right]_{\min} = \sum_j \frac{m^l}{4sb_j^2} = \frac{10^l - 1}{4} \quad (26)$$

This value depends on the number of qubits but does not depend on the type of polyhedron. It defines the minimum possible losses for considered protocols that do not use projections on entangled states.

Recall that in the general case for any protocols including those that project on entangled states minimum (optimal) losses are described by this equation:

$$L_{\min}^{opt} = \frac{v^2}{4(s-1)} = \frac{(2^l + 1)^2 (2^l - 1)}{4} \quad (27)$$

From comparison of the equations (26) and (27) we can see that these protocols provide minimum possible (optimal) losses for reconstruction of mixed states of full rank only for single-qubit states. We can again conclude that for multi-qubit cases protocols that provide minimum possible losses during quantum states reconstruction should necessarily include projections on entangled states.

## 10. Tests of the universal statistical distribution

On the next figure we present results of numerical experiments that test the universal statistical distribution for Fidelity. Two hundred experiments were conducted with sample size 1 million each.

The measurement protocol was based on tetrahedron. Four-qubit state that represents a mix of GHZ state and uniform density matrix (white noise) was tested. The density matrix for the state is:

$$\rho = f \frac{E}{16} + (1 - f) |GHZ\rangle\langle GHZ|, \quad (28)$$

where $E$ is the unit matrix of size 16 by 16, $|GHZ\rangle$ is the state of Greenberger- Horne- Zeilinger: $|GHZ\rangle = \frac{1}{\sqrt{2}} (|0000\rangle + |1111\rangle)$, $f$ is the weight of the uniform density matrix (white noise). In our

case $f = 0.5$ (50%). It is a multiparametric distribution. The size of vector of parameters is 255.

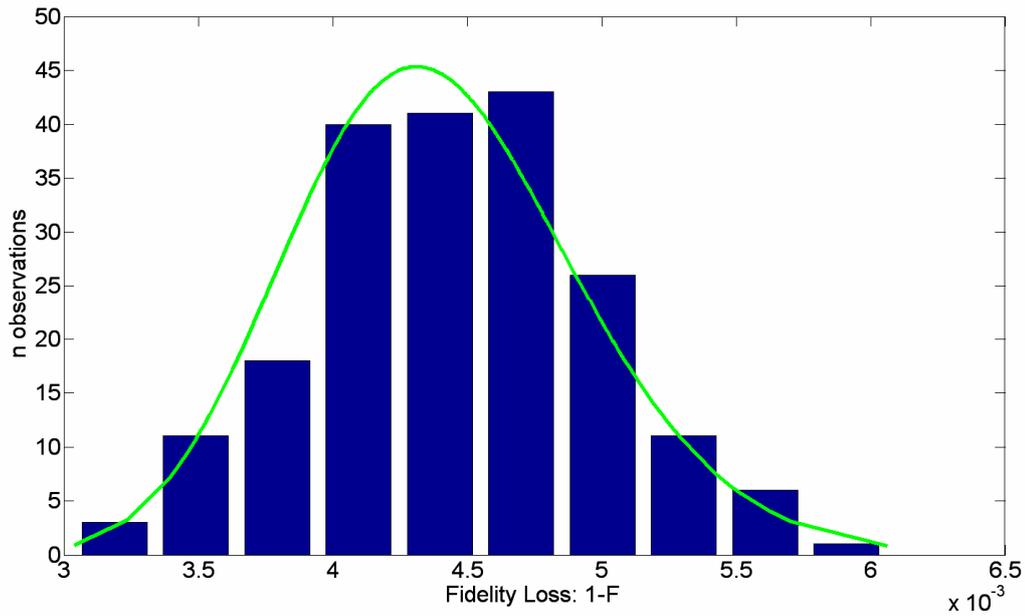

Figure 5. Fidelity loss distribution for 200 experiments

This data demonstrates good agreement between results of numerical experiment and theory developed in [34] with high critical significance level (0.65) for chi-squared criterion.

## 11. Dependence of reconstruction precision on the weight of "white noise"

The value of Fidelity can lie in a wide interval, so it is convenient to use a new variable $z = -\lg(1-F)$. The new variable $z$ defines the number of nines in Fidelity. For example $z = 3$ means that $F = 0.999$.

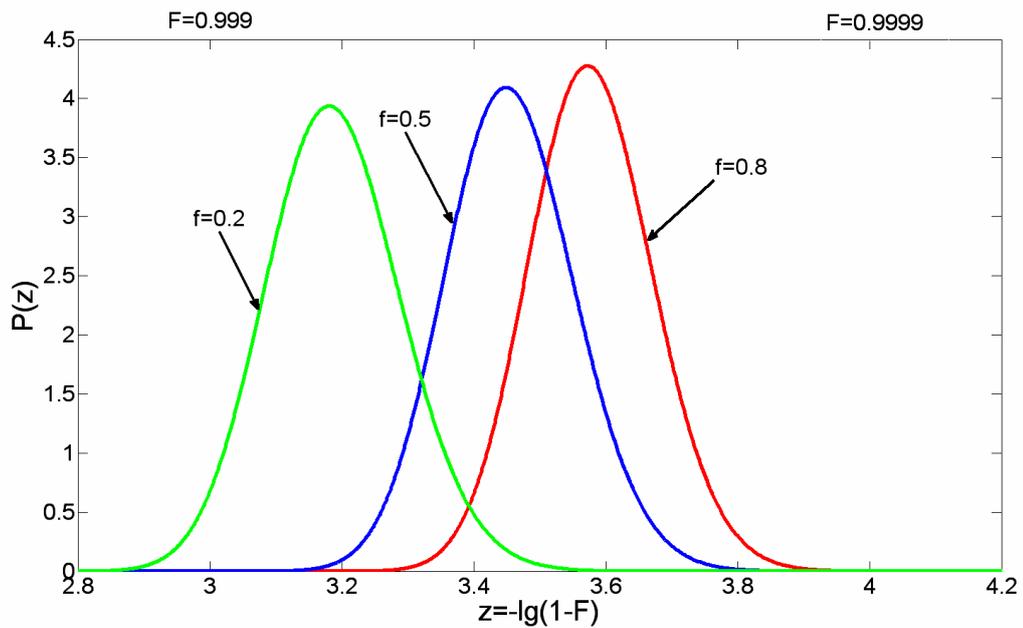

Figure 6. Theoretical fidelity distribution for three different states

The figure demonstrates distributions using the new variable for a three qubit state, which is again a mix of GHZ state and "white noise" (uniform density matrix). In our calculations we use a protocol based on dodecahedron. Sample size *n* is equal to one million. We demonstrate the dependence on the white noise weight. It is obvious that the higher the white noise weight the higher the precision of reconstruction. It is not too difficult to understand this fact in the context of our previous discussion about mixed states – "white noise" is best one for reconstruction.

## 12. Precision of reconstruction of Bell and GHZ states

On the next figure we demonstrate dependence of reconstruction of Bell and GHZ state on the number of qubits.

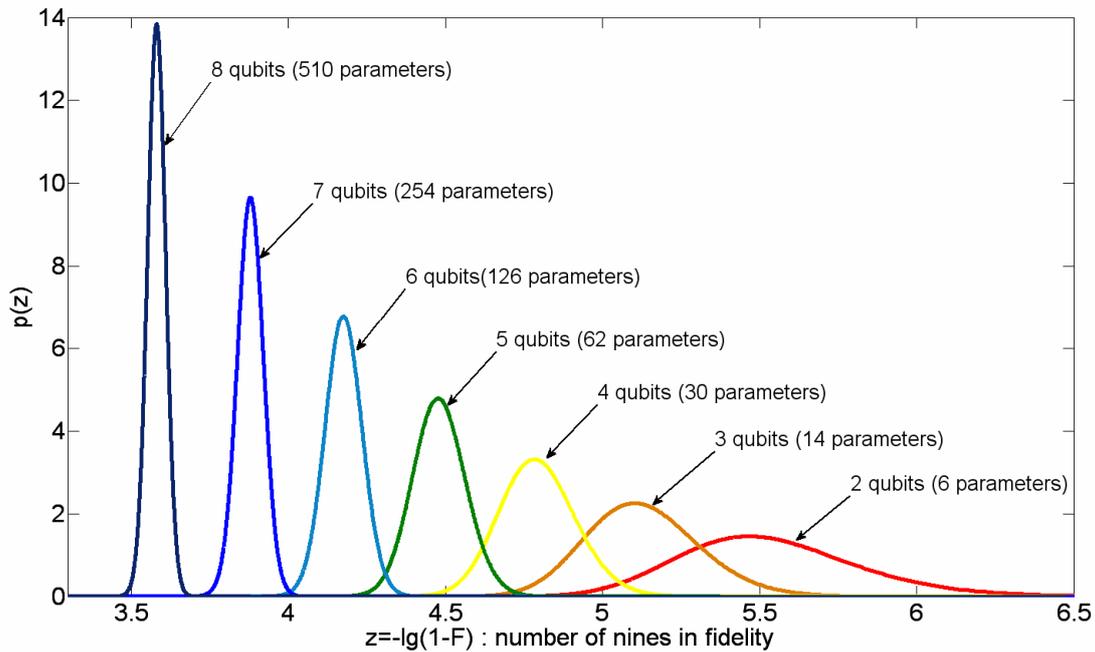

Figure 7. Theoretical fidelity distribution for Bell and GHZ states

We present registry with the number of qubits from two to eight. The sample size is one million. The measurement protocol was based on tetrahedron. With higher number of qubits the precision of reconstruction falls and the width of distribution falls too.

## 13. Test of adequacy of the model

Adequacy of quantum measurements means an internal agreement between statistical data and the theoretical model of quantum state. The test of adequacy is possible only if the protocol has some redundancy, i.e. when the number of rows is larger than the number of parameters to be estimated.

To test adequacy we can use chi-squared criterion:

$$\chi_r^2 = \sum_{j=1}^{m} \frac{(k_j - \lambda_j t_j)^2}{\lambda_j t_j} \quad (29)$$

Here $\lambda_j$ are estimates of events generation intensities that are obtained after solving the likelihood equation. If the model is adequate then this characteristic should have chi-squared distribution with the number of degrees of freedom equal to $v = m - (2s - r)r$ [34]. Results of numerical modeling which are presented on the figure above prove the validity of presented criterion.

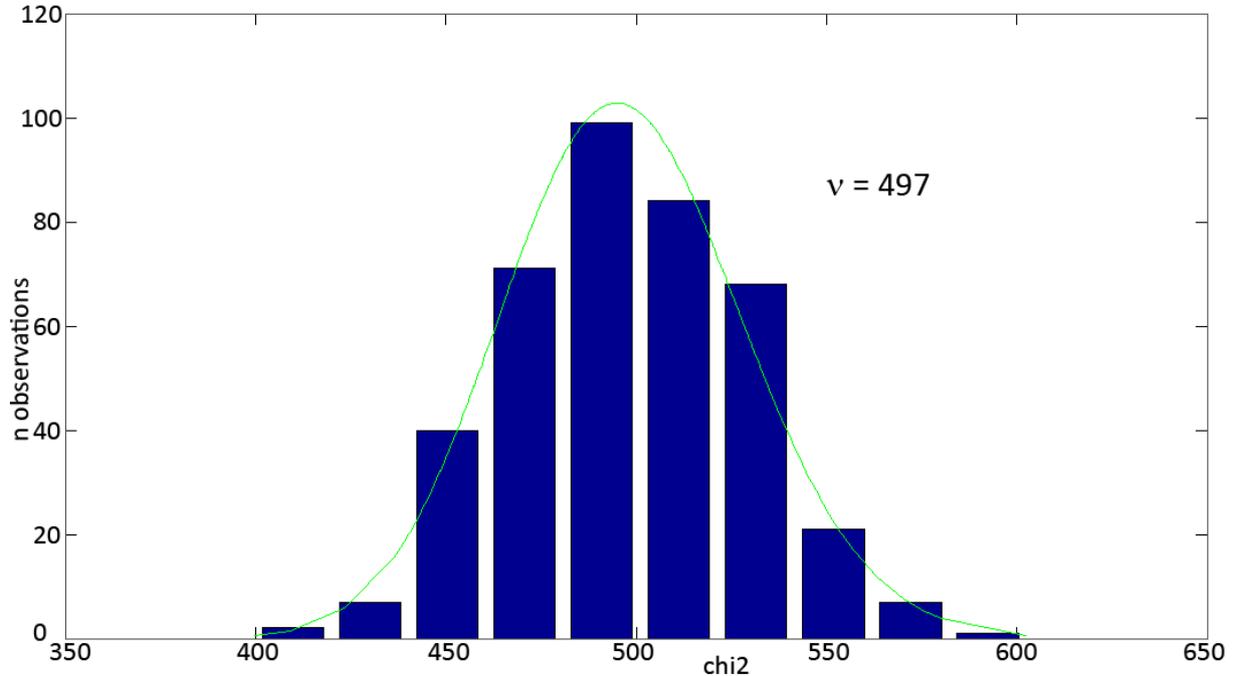

Figure 8. Test of adequacy in the case of adequate model

## 14. Conclusions

We consider a new methodology of quantum tomography protocols' quality estimation that is based on analysis of completeness, adequacy and precision of quantum measurements.

Our approach is based on analysis of a specially constructed measurement matrix and on the use of the universal statistical distribution for fidelity. Efficiency of the proposed approach is demonstrated for a large set of protocols and states.

This approach has important advantages. First of all, using purification procedure we can reconstruct quantum states in Hilbert spaces of rather high dimension. Secondly, using general statistical distribution for fidelity we can completely analyze precision of quantum tomography for any measurement protocols and states (both pure and mixed). Also, method allows the experimentator to use his resources in the most efficient way to construct an optimal measurement protocol.

The efficiency of this approach was demonstrated experimentally together with the group of Professor Sergey Kulik from Moscow State University and the group of Doctor Marco Genovese from INRIM Institute in Turin, Italy [32, 33].